\documentclass[prb,superscriptaddress,preprint,showpacs,floatfix]{revtex4}
\usepackage{amsmath}
\usepackage{amssymb}
\usepackage{graphicx}
\usepackage{hyperref}
\usepackage{natbib}

\newcommand{\eps}{\varepsilon}

\newcommand{\romand}{\mathrm{d}}  % normale Schrift (roman) fuer Integrale

\newcommand{\bfk}{\mathbf{k}}  % fett
\newcommand{\bfq}{\mathbf{q}}

\newcommand{\bfG}{\mathbf{G}}

\begin{document}
 \title{Physical and mathematical justification of the numerical Brillouin zone integration of the Boltzmann rate equation by Gaussian smearing}
 \author{Christian Illg}
 \author{Michael Haag}
 \affiliation{Max Planck Institute for Intelligent Systems, Heisenbergstr. 3, 70569 Stuttgart, Germany}
 \author{Nicolas Teeny}
 \affiliation{Max Planck Institute for Nuclear Physics, Saupfercheckweg 1, 69117 Heidelberg, Germany}
 \author{Jens Wirth}
 \affiliation{Institut f\"ur Analysis, Dynamik und Modellierung, Universit\"at Stuttgart, Pfaffenwaldring~57, 70569 Stuttgart, Germany}
 \author{Manfred F\"ahnle}
 \email[Electronic address: ]{faehnle@is.mpg.de}
 \affiliation{Max Planck Institute for Intelligent Systems, Heisenbergstr. 3, 70569 Stuttgart, Germany}

\begin{abstract}
Scatterings of electrons at quasiparticles or photons are very important for many topics in solid state physics, e.g., spintronics, magnonics or photonics, and therefore a correct numerical treatment of these scatterings is very important. For a quantum-mechanical description of these scatterings Fermi's golden rule is used in order to calculate the transition rate from an initial state to a final state in a first-order time-dependent perturbation theory. One can calculate the total transition rate from all initial states to all final states with Boltzmann rate equations involving Brillouin zone integrations. The numerical treatment of these integrations on a finite grid is often done via a replacement of the Dirac delta distribution by a Gaussian. The Dirac delta distribution appears in Fermi's golden rule where it describes the energy conservation among the interacting particles. Since the Dirac delta distribution is a not a function it is not clear from a mathematical point of view that this procedure is justified. We show with physical and mathematical arguments that this numerical procedure is in general correct, and we comment on critical points.
\end{abstract}

\pacs{02.60.Cb, 02.60.Jh, 02.60.Nm}

\maketitle

\section{Introduction}
\label{sec1}
In solid state physics scatterings of electrons at periodic perturbations (quasiparticles or photons) are very important for many research fields and we give three examples in the following: 
\begin{enumerate}
 \item In all-optical switching experiments \cite{Kirilyuk10} a thin ferrimagnetic film, e.g., GdFeCo, is irradiated by a femtosecond laser pulse which can be linearly or circularly polarized and thereafter a demagnetization with subsequent switching of the magnetization can be observed under certain preconditions. The fundamental mechanisms are strongly debated at the moment, however, electron-photon scatterings, electron-phonon scatterings and electron-magnon scatterings certainly play a big role for the demagnetization of the ferrimagnetic film.
 \item In ultrafast demagnetization experiments \cite{Koopmans10} a thin ferromagnetic film, e.g., Ni or Fe, is irradiated by a femtosecond laser pulse which is normally linearly polarized and thereafter an ultrafast demagnetization (on the time scale of about 100~femtoseconds) without switching of the magnetization can be observed. The magnetization recovers on a time scale of several picoseconds. Despite many years of research the fundamental mechanisms are still unclear but scatterings of electrons at phonons \cite{Essert11,Illg13} or at magnons \cite{Haag14} or at electrons \cite{Mueller13} have been discussed intensively.
 \item Spin-polarized currents are important for devices in spintronics \cite{Zutic04}, e.g., spin-transistors or spin-diodes. The lifetime of the spin-polarized electrons is crucial for the spintronics devices. The lifetimes are determined by scatterings of electrons at quasiparticles and at interfaces or defects.
\end{enumerate}
A correct numerical calculation of the various scattering processes is important for the understanding of these effects in solid state physics. In quantum mechanics Fermi's golden rule gives the transition rate $W_{j\bfk, j'\bfk'}^\lambda$ from an initial electronic state $\Psi_{j\bfk}$ in a solid with energy $\eps_{j\bfk}$ to a final electronic state $\Psi_{j'\bfk'}$ with energy $\eps_{j'\bfk'}$ ($j$,$j'$: band indices; $\bfk$, $\bfk'$: wavevectors) due to a periodic perturbation arising from a (quasi)particle \cite{Faehnle11}
\begin{align}
 W_{j\bfk, j'\bfk'}^\lambda =& \frac{2\pi}{\hbar} 
\left| M_{j\bfk ,j'\bfk'}^\lambda \right|^2 \cdot \delta\left( \eps_{j'\bfk'} - ( \eps_{j\bfk} \pm \hbar \omega_{\bfq\lambda} ) \right).
\label{eq1.1}
\end{align}
$\pm \hbar \omega_{\bfq\lambda}$ is the energy of the involved (quasi)particle ($\bfq$: wavevector, $\lambda$: polarization) which may be, e.g., photons, phonons, magnons, plasmons etc.\ with frequency $\omega_{\bfq\lambda}$ for absorption (plus sign) or emission (minus sign), and $M_{j\bfk ,j'\bfk'}^\lambda$ is the scattering matrix element
\begin{align}
 M_{j\bfk,j'\bfk'}^\lambda = \langle F' \Psi_{j'\bfk'} \left| W_{\bfq\lambda} \right| F \Psi_{j\bfk } \rangle ,
\label{eq1.2}
\end{align}
where $\left| F \right. \rangle$ and $\left| F' \right. \rangle$ are the initial and final (quasi)particle states and $W_{\bfq\lambda}$ is the scattering operator. Thereby, momentum conservation $\bfk\pm\bfq=\bfk'+\bfG$ is demanded ($\bfG$: reciprocal lattice vector). Fermi's golden rule is the first-order approximation of the time-dependent quantum-mechanical perturbation theory. It implies that the scattering processes are Markovian which means that a scattering process does not depend on preceding scattering processes. Fermi's golden rule is only valid in a time window where the perturbation time on the one hand must be short enough because of the first-order approximation and on the other hand must be long enough in order to replace the $\sin(x)/x$-function appearing in the derivation of Fermi's golden rule by the Dirac delta distribution. The validity of Fermi's golden rule for a magnetization dynamics on the 100~fs timescale is critically discussed in Ref.~\onlinecite{Illg13}.

Normally, one is not interested in a specific transition rate $W_{j\bfk, j'\bfk'}^\lambda$ from an initial state $\Psi_{j\bfk}$ to a final state $\Psi_{j'\bfk'}$ but in the total transition rate $W_\mathrm{total}$ from all initial states to all final states. 
Thereby $\mathbf{k}$ and $\mathbf{k}'$ are related via $\mathbf{k}\pm\mathbf{q}=\mathbf{k}'+\mathbf{G}$ if the scattering
is at a quasiparticle with wavevector $\mathbf{q}$.
This is calculated with Boltzmann rate equations \cite{Illg13,Yafet63}
\begin{align}
 W_\mathrm{total} = W_\mathrm{in} - W_\mathrm{out}
\label{eq1.3}
\end{align}
where
\begin{align}
W_\mathrm{in}= & \frac{1}{\Omega_\mathrm{BZ}^2}\ \sum_{j,j',\lambda}  \int_\mathrm{BZ}\romand^3 k \int_\mathrm{BZ}\romand^3 k' \ n_{j' \bfk'}  \left[ 1-n_{j \bfk} \right]  W_{j'\bfk',j\bfk}^{\lambda}
\label{eq1.4}\\
W_\mathrm{out}= & \frac{1}{\Omega_\mathrm{BZ}^2}\ \sum_{j,j',\lambda}  \int_\mathrm{BZ}\romand^3 k \int_\mathrm{BZ}\romand^3 k' \ n_{j \bfk}  \left[ 1-n_{j' \bfk'} \right]  W_{j\bfk,j'\bfk'}^{\lambda}.
\label{eq1.5}
\end{align}
$\Omega_\mathrm{BZ}$ is the Brillouin zone (BZ) volume and $n$ is the distribution function for the electrons.

Often one is also interested in the rate of change of the distribution function $n_{j\bfk}$ due to scattering which is also calculated with Boltzmann rate equations \cite{Ashcroft76}
\begin{align} 
  \frac{\romand n_{j\bfk}}{\romand t} =&  \frac{1}{\Omega_\mathrm{BZ}}\ \sum_{j',\lambda}  \int_\mathrm{BZ}\romand^3 k'\ \Big\{  n_{j' \bfk'}  \left[ 1-n_{j\bfk} \right]  W_{j'\bfk',j\bfk}^{\lambda} - 
n_{j \bfk}  \left[1-n_{j' \bfk'} \right]  W_{j\bfk,j'\bfk'}^{\lambda} \Big\}.
\label{eq1.7}
\end{align}
So we have to calculate Brillouin zone integrals of the form
\begin{align}
 \int_\mathrm{BZ} \romand^3 k\ g(\bfk)\ \delta(\eps(\bfk)).
\label{eq1.7a}
\end{align}
Because the quantities $\eps_{j\bfk}$, $\eps_{j'\bfk'}$, $W_{j'\bfk',j\bfk}^{\lambda}$, $W_{j\bfk,j'\bfk'}^{\lambda}$ can be calculated numerically only for a finite number of $\bfk$-points, finite $\bfk$-point grids have to be used for the numerical calculation of the total transition rate $W_\mathrm{total}$ or of the rate of change of the distribution function $\romand n_{j\bfk}/\romand t$. Thereby, energy conservation $\eps_{j'\bfk'} = \eps_{j\bfk} \pm \hbar \omega_{\bfq\lambda}$ and momentum conservation $\bfk\pm\bfq=\bfk'+\bfG$ have to be fulfilled, however, energy conservation in combination with momentum conservation is in general never fulfilled for a finite $\bfk$-point grid. Therefore, the Dirac delta distribution has to be replaced by a ``smeared'' delta function in order to obtain a result which approximates the integral (which is done, e.g., in Refs. \cite{Essert11,Illg13,AlShaikhi07,franchini,Paulatto} and in very many other papers)
. To do this, often the following equation is used 
\begin{align}
 \int_\mathrm{BZ} \romand^3 k\ g(\bfk)\ \delta(\eps(\bfk)) \approx \int_\mathrm{BZ} \romand^3 k\ g(\bfk)\ \frac{1}{\sqrt{\pi} \sigma} \exp\left(-\frac{\eps^2(\bfk)}{\sigma^2}\right)
\label{eq1.8}
\end{align}
and the smearing parameter $\sigma$ has to be chosen appropriately, see Sec.~\ref{sec3}. This means that the contribution of a certain grid point to the total transition rate $W_\mathrm{total}$ or to the rate of change of the distribution function $\romand n_{j\bfk}/\romand t$ is small if the energy conservation is fulfilled very badly, and vice versa the contribution is large if the energy conservation is fulfilled very well. However, from a mathematical point of view it is not obvious that Eq.~\eqref{eq1.8} holds since the Dirac delta distribution is not a function and the smearing is with respect to the energy $\eps$ but the integration is with respect to the wavevector $\bfk$. The problem is explained in more detail in Sec.~\ref{sec2}. 

Mathematical proofs of Eq.~\eqref{eq1.8} under certain preconditions can be found in Ref.~\onlinecite{Friedlander98}, theorem 7.2.1, and in Ref.~\onlinecite{Hoermander98}, theorem 6.1.5, however, the proofs are for general distributions and are very abstract. We want to show in this article that Eq.~\eqref{eq1.8} is correct by using also physical arguments. 

%In Sec.~\ref{sec2} we give a short overview about the numerical integration of integrals involving the Dirac delta distribution, $\delta( \eps(\bfk))$, and show that the above described smearing is correct for $\delta(\eps(\bfk))$. 

 In Sec.~\ref{sec2} we explain in detail the problem which arises when in a Brillouin-zone integration
 Dirac's delta distribution is approximated by a Gaussian.
 This is done in many papers without giving any justification.
 We therefore think that the outline of this problem is  a novelty per se.
 Then we give a justification of the Gaussian smearing method by mathematical and physical arguments.
 Each of these arguments has been used in other contexts in previous papers.
 The novelty of our paper is that we use these arguments to  justify the Gaussian smearing method for the Brillouin-zone integration.
 First, we consider a coordinate transformation from the wavevector variables $\mathbf{k}=\left(k_{x},k_{y},k_{z}\right)$
 to the variables $\epsilon,\vartheta,\varphi$ where $\epsilon$ is the energy and $\vartheta,\varphi$ are variables for the surface
 of constant energy.
 This transformation involves the Jacobian $J\left( k_{x},k_{y},k_{z}\right)$.
 The inverse function theorem \cite{Friedlander98} says that if this Jacobian is nonzero at a $\mathbf{k}$-point,
 this transformation is invertible.
 Then the integration in $\mathbf{k}$-space including $\delta{\left(\epsilon\right)}$ can be represented by an integration over $\epsilon,\vartheta,\varphi$ of a function which involves the Jacobian   $\widetilde{J}\left(\epsilon,\vartheta,\varphi\right)=
 \left[J\left(k_{x},k_{y},k_{z}\right)\right]^{-1}$ and,
 which now can without any problem be replaced by a Gaussian.
 The problem is that there are special $\mathbf{k}$-points where $\nabla_{\mathbf{k}}\epsilon\left(\mathbf{k}\right)=0$.
 For these special points the Jacobian $J\left(k_{x},k_{y},k_{z}\right)$ is zero,
 and the reverse transformation involving $\widetilde{J}\left(\epsilon,\vartheta,\varphi\right)$ 
 is not defined in a rigorous mathematical interpretation.
 According to a general  theorem of M. Morse the dispersion relation $\epsilon\left(\mathbf{k}\right)$
 exhibits such special points because it is periodic in all components.
 There are special points which can be identified easily, e.g., the $\Gamma$-point and points on the Brillouin zone boundary.
 These points can be avoided by shifting the grid of $\mathbf{k}$-points
 considered in the Brillouin zone integration 
 accordingly \cite{Martinbook}. Other special points cannot be easily found,
 and they might be in the shifted k-point grid.
 Van Hove has shown \cite{vanHove} that for three dimensions the appearance of these 
 special points does not appreciably modify the result of a numerical integration.
 We motivate these steps by physical reasoning.

In Sec.~\ref{sec3} we give practical hints for the appropriate choice of the smearing parameter $\sigma$. Finally, our results are summarized in Sec.~\ref{sec4}.

\section{Numerical integration of the Dirac delta distribution}
\label{sec2}
It is very well known that in integrals involving the Dirac delta distribution the distribution can be replaced by a Gaussian for the limes $\sigma \rightarrow 0$. It reads
\begin{align}
 &\int_{-\infty}^{+\infty} \romand \eps\ g(\eps)\ \delta(\eps) = \lim_{\sigma\rightarrow 0}  \int_{-\infty}^{+\infty} \romand \eps\ g(\eps)\ \frac{1}{\sqrt{\pi} \sigma} \exp\left(-\frac{\eps^2}{\sigma^2}\right)
\label{eq2.1}
\end{align}
where $g(\eps)$ is a continuously differentiable function which depends on the energy $\eps$. The Dirac delta distribution is approximated by a Gaussian
\begin{align}
 &\int_{-\infty}^{+\infty} \romand \eps\ g(\eps)\ \delta(\eps) \approx  \int_{-\infty}^{+\infty} \romand \eps\ g(\eps)\ \frac{1}{\sqrt{\pi} \sigma} \exp\left(-\frac{\eps^2}{\sigma^2}\right)
\label{eq2.2}
\end{align}
for a numerical calculation of the integral and $\sigma$ has to be chosen appropriately, see Sec.~\ref{sec3}. 

However, the integrals in Eqs.~\eqref{eq1.3}, \eqref{eq1.4}, \eqref{eq1.5} and \eqref{eq1.7} are not over the energy $\eps$ but over the wavevector $\bfk$. For the sake of simplicity we discuss the following integral  
\begin{align}
 \int_\mathrm{BZ} \romand^3 k\ g(\bfk)\ \delta(\eps(\bfk))
\label{eq2.3}
\end{align}
where $g(\bfk)$ is a continuously differentiable function of the wavevector $\bfk$, and the generalization to the expression $\delta\left( \eps_{j'\bfk'} - ( \eps_{j\bfk} \pm \hbar \omega_{\bfq\lambda} ) \right)$ used in Eqs.~\eqref{eq1.3}, \eqref{eq1.4}, \eqref{eq1.5} and \eqref{eq1.7} is straightforward. In explicit numerical calculations it is always assumed that also the relation
\begin{align}
 \int_\mathrm{BZ} \romand^3 k\ g(\bfk)\ \delta(\eps(\bfk)) = \lim_{\sigma\rightarrow 0} \int_\mathrm{BZ} \romand^3 k\ g(\bfk)\ \frac{1}{\sqrt{\pi} \sigma} \exp\left(-\frac{\eps^2(\bfk)}{\sigma^2}\right)
\label{eq2.4a}
\end{align}
holds without giving any justification, reference or comment and that this may be approximated by
\begin{align}
 \int_\mathrm{BZ} \romand^3 k\ g(\bfk)\ \delta(\eps(\bfk)) \approx \int_\mathrm{BZ} \romand^3 k\ g(\bfk)\ \frac{1}{\sqrt{\pi} \sigma} \exp\left(-\frac{\eps^2(\bfk)}{\sigma^2}\right).
\label{eq2.4}
\end{align}
However, the Dirac delta distribution is defined by Eq.~\eqref{eq2.1} and not by Eq.~\eqref{eq2.4a}. We show in the following how the use of  Eq.~\eqref{eq2.4} can be justified. 

We consider a coordinate transformation from the wavevector variables $\bfk=(k_x, k_y, k_z)$ to the variables $\eps, \vartheta, \varphi$ ($\eps$: energy; $\vartheta,\varphi$: variables for the surface of constant energy)
\begin{align}
 \eps &= \eps(k_x, k_y, k_z) \notag \\
 \vartheta &= \vartheta(k_x, k_y, k_z)  \notag \\
 \varphi &= \varphi(k_x, k_y, k_z) 
\label{eq2.7}
\end{align}
where the energy dispersion relation $\eps(k_x, k_y, k_z)$ is known and the surfaces of constant energy can be parametrized with two variables $\vartheta$ and $\varphi$. The inverse function theorem says \cite{Spivak71} that every continuously differentiable, vector-valued function which maps values from an open set of $\mathbb{R}^n$ to other values of an open set of $\mathbb{R}^n$ (so-called coordinate transformation, e.g., Eq.~\eqref{eq2.7}) and whose Jacobian determinant
\begin{align}
J(k_x,k_y,k_z) = 
\det \left( \begin{array}{ccc}
\frac{\partial \eps}{\partial k_x} & \frac{\partial \eps}{\partial k_y} & \frac{\partial \eps}{\partial k_z} \\
\frac{\partial \vartheta}{\partial k_x} & \frac{\partial \vartheta}{\partial k_y} & \frac{\partial \vartheta}{\partial k_z} \\
\frac{\partial \varphi}{\partial k_x} & \frac{\partial \varphi}{\partial k_y} & \frac{\partial \varphi}{\partial k_z} \end{array} \right)
\label{eq2.6}
\end{align}
is non-zero at a point is invertible in the neighborhood of this point, i.e., the reverse transformation of Eq.~\eqref{eq2.7} 
\begin{align}
 k_x &= k_x(\eps, \vartheta, \varphi)  \notag \\
 k_y &= k_y(\eps, \vartheta, \varphi)  \notag \\
 k_z &= k_z(\eps, \vartheta, \varphi)
\label{eq2.8}
\end{align}
exists and can in principle be given in the neighborhood of every point $(k_x, k_y, k_z)$ if the above-mentioned conditions are fulfilled. 

If Eq.~\eqref{eq2.7} is invertible in the neighborhood of every point $\bfk=(k_x, k_y, k_z)$---whereby only points $\bfk$ with $\eps(\bfk)=0$ are relevant because of the Dirac delta distribution in Eq.~\eqref{eq2.3}---, it is possible to make for this neighborhood a local coordinate transformation (using Eq.~\eqref{eq2.8}) for the function $g(\bfk)=\widetilde{g}(\eps,\vartheta,\varphi)$ appearing in Eq.~\eqref{eq2.3}. Then, the integral over the wavevector $\bfk$ can be replaced by the integral over the variables $\eps,\vartheta,\varphi$
\begin{align} 
 & \int_\mathrm{BZ} \romand^3 k\ g(\bfk)\ \delta(\eps(\bfk))  \notag \\
 & = \int \romand \eps\ \int \romand \vartheta\ \int \romand \varphi\ |\widetilde{J}(\eps,\vartheta,\varphi)| \cdot g\big(k_x(\eps,\vartheta,\varphi),k_y(\eps,\vartheta,\varphi),k_z(\eps,\vartheta,\varphi)\big) \cdot \delta(\eps)  \notag \\
& = \int \romand \eps\ \int \romand \vartheta\ \int \romand \varphi\ |\widetilde{J}(\eps,\vartheta,\varphi)| \cdot \widetilde{g}(\eps,\vartheta,\varphi) \cdot \delta(\eps) 
\label{eq2.5}
\end{align}
where $\widetilde{J}(\eps,\vartheta,\varphi)$ is the Jacobian determinant of the reverse transformation~\eqref{eq2.8}
\begin{align}
\widetilde{J}(\eps,\vartheta,\varphi) = 
\det \left( \begin{array}{ccc}
\frac{\partial k_x}{\partial \eps} & \frac{\partial k_x}{\partial \vartheta} & \frac{\partial k_x}{\partial \varphi} \\
\frac{\partial k_y}{\partial \eps} & \frac{\partial k_y}{\partial \vartheta} & \frac{\partial k_y}{\partial \varphi} \\
\frac{\partial k_z}{\partial \eps} & \frac{\partial k_z}{\partial \vartheta} & \frac{\partial k_z}{\partial \varphi} \end{array} \right).
\label{eq2.6b}
\end{align}
Note that $\widetilde{J}(\eps,\vartheta,\varphi)=J^{-1}(k_x,k_y,k_z)$ with $(\eps, \vartheta, \varphi)$ expressed by Eq.~\eqref{eq2.7}. It is now definitely allowed to approximate the Dirac delta distribution by a Gaussian in analogy to Eqs.~\eqref{eq2.1} and \eqref{eq2.2}
\begin{align}
 & \int_\mathrm{BZ} \romand^3 k\ g(\bfk)\ \delta(\eps(\bfk))  \notag \\
 &\approx 
\int \romand \eps\ \int \romand \vartheta\ \int \romand \varphi\ |\widetilde{J}(\eps,\vartheta,\varphi)| \cdot \widetilde{g}(\eps,\vartheta,\varphi) \cdot \frac{1}{\sqrt{\pi} \sigma } \exp\left(-\frac{\eps^2}{\sigma^2}\right) \notag \\
 &= \int \romand^3 k\ g(\bfk) \frac{1}{\sqrt{\pi} \sigma} \exp\left(-\frac{\eps^2(\bfk)}{\sigma^2}\right)
\label{eq2.11}
\end{align}
where in the last step the integration variables are changed back again to an integration over the wavevector using Eq.~\eqref{eq2.7}. This is exactly what we wanted to show in Eq.~\eqref{eq2.4}.

One must keep in mind that the Jacobian determinant $J(k_x,k_y,k_z)$ given by Eq.~\eqref{eq2.6} is zero for special points $\bfk=(k_x, k_y, k_z)$ where $\nabla_\bfk \eps(\bfk)=0$. This is the case for the $\Gamma$-point and usually for points on the Brillouin zone boundary \cite{Haug1}, and even the transformation~\eqref{eq2.7} could be not continuously differentiable for special points. Then, the reverse transformation~\eqref{eq2.8} used in Eq.~\eqref{eq2.11} is not defined anymore in a rigorous mathematical interpretation. However, these problems arise because of two idealizations, the long-time idealization and the infinite-solid idealization, and the following remarks have to be considered:
\begin{enumerate}
 \item In a physical interpretation the Dirac delta distribution appearing in Fermi's golden rule, Eq.~\eqref{eq1.1}, is only a long-time idealization which should be replaced by a $\sin(x)/x$-function for realistic physical calculations. However, this would yield time-dependent rates which is usually not desired.
 \item For a numerical calculation an infinite periodicity of the lattice is assumed (infinite-solid idealization). $\bfk$-points of this numerical calculation only sometimes coincide exactly with a point where the reverse transformation~\eqref{eq2.8} is not defined and the $\bfk$-point grid can always be shifted so that there is no point where the reverse transformation is not defined. For an arbitrary $g(\bfk)$ it is not clear that one gets a correct result when omitting these $\bfk$-points. In a real solid in the ground state only a finite number of energy levels are occupied. These energy levels do not correspond to states with defined wavevectors $\bfk$. In the numerical treatment of these finite systems the energy levels are approximated by the energies of a lattice with infinite periodicity at a number of discrete points on a $\bfk$-point grid. For sufficiently large systems the result must be independent of the detailed choice of the $\bfk$-point grid. If we know the critical points, we can shift the 
$\bfk$-point grid in such a way that it does not include the above defined critical points (see also Ref.\ \onlinecite{Martinbook}).
One could argue that a $\bfk$-point very close to a critical $\bfk$-point could yield an extremely large contribution because the Jacobian determinant $\widetilde{J}(\eps,\vartheta,\varphi)$~\eqref{eq2.6b} appearing in Eq.~\eqref{eq2.5} may be very large for this point. This means that the second line of Eq.~\eqref{eq2.11} would not be a good approximation for a choice of the $\bfk$-point grid which contains points very close to critical points, but the near-equality of the first line with the third line of Eq.~\eqref{eq2.11} still holds because for the transition between the first and the third line the integration variables have been changed back again and this corresponds to the multiplication with $\widetilde{J}^{-1}(\eps,\vartheta,\varphi)=J(k_x,k_y,k_z)$, so altogether, a possibly large value of $\widetilde{J}(\eps,\vartheta,\varphi)$ does not matter. However, it is extremely complicated to identify all critical 
points and therefore it is not clear whether a chosen $\bfk$-point grid contains critical points or not. In order to avoid these cumbersome investigations one can also do the following: One performs calculations for denser and denser grids and/or for shifted and rotated grids and compares the results. If the results are very similar, this means that the grids either do not contain critical points or that the critical points do not make a big contribution so that they do not falsify the results, in agreement with Ref.\ \onlinecite{vanHove}.  
\end{enumerate}

\section{Practical hints for the appropriate choice of the smearing parameter}
\label{sec3}
The appropriate choice of the smearing parameter $\sigma$ appearing in Eq.~\ref{eq2.2} is crucial for the correct numerical calculation of the Boltzmann rate equation. The appropriate choice of $\sigma$ depends on two quantities:
\begin{enumerate}
 \item First, the smearing parameter $\sigma$ depends on the energy scale of the involved (quasi)particle which may be, e.g., a photon, phonon, magnon, plasmon (see Sec.~\ref{sec1}). The energy scale for a phonon is in the order of some mRy (about 40~meV) and for a magnon the energy scale is a factor 10 larger than for a phonon. The energy scale for a plasmon is much larger, in the order of 700~mRy (about 10~eV). An appropriate choice of the smearing parameter is in the same order as the energy scale of the involved (quasi)particle.
 \item Second, the smearing parameter $\sigma$ depends on the grid spacing. A typical ansatz is $\sigma= p / N_1$ where $N_1$ is the number of $\bfk$-points in one direction and the total number of $\bfk$-points is $N_1^3$. For a fixed proportionality constant $p$ it is guaranteed that the smearing parameter is the smaller the larger $N_1$. 
\end{enumerate}
In the following we discuss the choice of $\sigma$ for the case of electron-phonon scatterings. In Ref.~\onlinecite{Essert11} the smearing parameter is fixed to 15~meV (about 1~mRy) for the numerical calculation of the electron-phonon Boltzmann rate equation. In Ref.~\onlinecite{Illg13} we tested our numerical results of the electron-phonon Boltzmann rate equation for many different grids and smearing parameters. In this publication we considered the case of ultrafast demagnetization, see Sec.~\ref{sec1}. Among other quantities we calculated the rate of the magnetic moment change per atom $dM/dt(t_s)$ for a time $t_s$ (see Eq.~(14) of Ref.~\onlinecite{Illg13}). $t_s$ is the time after the laser pulse irradiation where the electron system has thermalized, i.e., the electron distribution can be described by a Fermi-Dirac distribution with the electron temperature $T_e$.
\begin{figure}[]
 \centering
 \includegraphics[width=12.5cm]{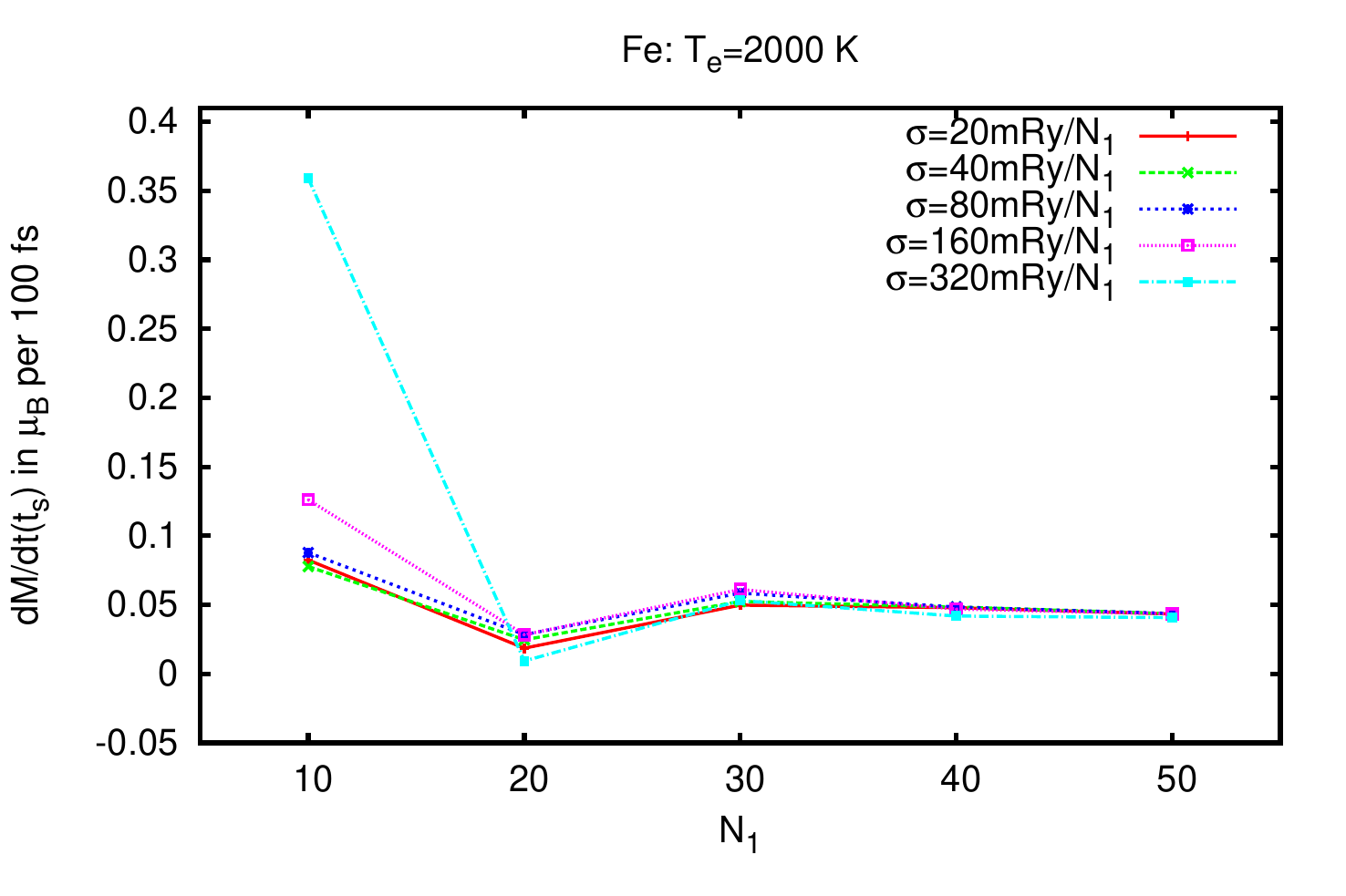}
 \caption{Rate of the magnetic moment change per atom $dM/dt(t_s)$ vs.\ number of k-points in one direction $N_1$ for different smearing parameters $\sigma$. For iron and an excitation temperature of $T_e=2000$~K.}
 \label{convdMdtFe}
\end{figure}
Fig.~\ref{convdMdtFe} shows the rate of the magnetic moment change per atom $dM/dt(t_s)$ for Fe and an electron temperature of $T_e=2000$~K. We calculated $dM/dt(t_s)$ for different grids (number of k-points in one direction $N_1$) and for different smearing parameters $\sigma$. One can see that the results for $dM/dt(t_s)$ depend hardly on the chosen grid and on the chosen smearing parameter except for $N_1=10$. Therefore, the above discussed critical points do not falsify our results and the smearing parameter is in the right order of magnitude (about several mRy).
Of course it is trivial that increasing the number of $\mathbf{k}$-points increases the convergence.
By Fig.\ \ref{convdMdtFe} we just want to show that the results depend  only very slightly on the specific choice of $\sigma$
if the number of $\mathbf{k}$-points is above a certain value.

\section{Conclusions}
\label{sec4}
Scattering processes of electrons at periodic perturbations are very important in solid state physics and also a correct numerical treatment is crucial for the quantitative analysis of scattering processes in many research activities, e.g., spintronics. In quantum mechanics Fermi's golden rule is used which contains the Dirac delta distribution. The Dirac delta distribution is usually replaced by a Gaussian in order to integrate numerically the Boltzmann rate equations on a finite grid of $\bfk$-points. It is not obvious from the very beginning that this numerical treatment is correct since the Dirac delta distribution is not a function and the smearing variable differs from the integration variable. We have shown in the present article that this procedure is in general correct. There are special $\bfk$-points for which it is in principle not justified to replace the Dirac delta distribution by a Gaussian, however we have given mathematical and physical arguments why this procedure is nevertheless a good approximation for the integration of the Boltzmann rate equation and should not falsify the results,
at least for three dimensions. It is not clear whether the same holds also for d=2 or even for d=1.
In conclusion, the naive replacement of the Dirac delta distribution by a Gaussian gives in general correct results for the Boltzmann rate equation but this has to be checked for denser and/or for shifted and rotated grids in order to avoid wrong contributions from the above described special $\bfk$-points where the replacement is critical.

\bibliographystyle{apsrev4-1}
\bibliography{Submission} 

\end{document}